%% file: source.tex
\documentclass[pra,twocolumn,superscriptaddress,10pt]{revtex4-2}
\usepackage{float,graphicx,multirow, amsmath,color,dsfont}
\usepackage{amsmath,bm}
\usepackage{amssymb,mathrsfs,dsfont}
\usepackage{amssymb,mathbbol}
\usepackage{amsfonts}
\usepackage{mathrsfs}
\usepackage{float}
\usepackage{xparse}
\usepackage{subcaption}

\usepackage{xcolor,hyperref}
\definecolor{darkblue}{rgb}{0,0.0.1,0.3}
\definecolor{darkred}{rgb}{0.6,0.1,0}
\hypersetup{colorlinks,breaklinks,
	linkcolor=darkred,urlcolor=darkblue,
	anchorcolor=darkred,citecolor=
	darkred,pdfauthor=RCB, pdftitle=RCB.Phase}
\usepackage{tikz}
\usepackage{mathbbol}
\usepackage{float}
\usepackage[normalem]{ulem}

\begin{document}	
	
\title{Husimi phase distribution in  non-Gaussian operations }
\author{Ramniwas Meena}
\email{meena.53@iitj.ac.in}
\affiliation{Department of Physics, Indian Institute of Technology Jodhpur, Rajasthan 342030, India.}

\author{Chandan Kumar}
\email{chandan.quantum@gmail.com}
\affiliation{Optics and Quantum Information Group, The Institute of Mathematical Sciences, CIT Campus, Taramani, Chennai 600113, India.}
\affiliation{Homi Bhabha National Institute, Training School Complex, Anushakti Nagar, Mumbai 400085, India.}

\author{Subhashish Banerjee}
\email{subhashish@iitj.ac.in}
\affiliation{Department of Physics, Indian Institute of Technology Jodhpur, Rajasthan 342030, India.}
	\begin{abstract}
	The Husimi phase distribution, an experimentally measurable quantity, is investigated for single-mode and two-mode squeezed vacuum states. The analysis highlights that non-Gaussian operations, i.e., photon subtraction (PS), photon addition (PA) and photon catalysis (PC), are effective tools for localizing phase distribution and enhancing phase robustness in the presence of noise, while PC enhances phase sensitivity but leads to greater delocalization. The work highlights the perspective that combined effects of squeezing, beam splitter transmittance, and environmental interactions must be carefully considered when quantum state engineering protocols are designed and phase properties provide a valuable insight into this endeavour.
    
	\end{abstract}
	\maketitle

    

\section{Introduction} 
The study of phase distributions in quantum optics and quantum information science is essential for understanding coherence properties, nonclassicality, and interference effects of quantum states ~\cite{SCHNABEL20171, perinova1998phase}. Unlike in classical wave theory, where phase is well-defined, the concept of phase in quantum mechanics is more intricate due to the uncertainty principle and the lack of a universally accepted phase operator ~\cite{DEMKOWICZDOBRZANSKI2015345}. As a result, various approaches have been developed to characterize quantum phase, including phase operators, direct phase measurement techniques, and phase-space-based probability distributions \cite{leonhardt1995canonical, agarwal1992classical}.

The canonical phase distribution is an idealized phase representation, constructed based on the principle that phase should be conjugate to the photon number operator. This distribution, derived using the Pegg-Barnett formalism \cite{pegg1989phase}, maintains strict phase-number complementarity, making it an important theoretical benchmark. It provides a fundamental description of phase independent of measurement processes, ensures phase-number complementarity, and serves as a reference model for understanding experimental deviations. However, it is not directly measurable due to inherent uncertainties in quantum phase measurement, and in low-photon-number regimes, it significantly deviates from experimentally obtained phase distributions \cite{luis1996optimum}.

The measured phase distribution corresponds to experimentally observed phase probabilities, incorporating noise and uncertainties introduced by detection techniques. Measurement strategies include optical homodyne tomography, heterodyne detection, and direct phase measurement techniques such as the Noh-Fougeres-Mandel (NFM) experiment \cite{noh1991measurement}. Measured phase distributions are effectively noisy versions of the canonical phase distribution, and in the semiclassical regime with high photon numbers, they approximate canonical phase distributions. However, they tend to be broader than canonical phase distributions due to measurement noise \cite{leonhardt1997measuring}. 

While phase operator-based approaches provide a fundamental theoretical framework, they suffer from measurement difficulties and inconsistencies in low-photon-number regimes ~\cite{pegg1989phase, luis1996optimum}. Measured phase distributions, obtained through homodyne tomography and heterodyne detection, are experimentally accessible but are broadened due to detection noise ~\cite{noh1991measurement, leonhardt1997measuring}. Phase-space-based representations, such as the Wigner function, Husimi function, and Glauber $P$-distribution, offer alternative approaches that incorporate quantum uncertainties while maintaining a closer analogy to classical phase distributions ~\cite{vogel1989determination, hudson1974wigner, moyal1949quantum, cohen1966generalized,lee1995theory}. They provide a realistic description of quantum optical phase properties, play a crucial role in phase estimation and quantum state reconstruction, and aid in analyzing quantum noise and decoherence effects in quantum technologies \cite{schleich1987oscillations,chizhov1993phase}.

Among these, the Husimi function is particularly well-suited for phase probability distribution analysis due to its non-negative nature and ability to incorporate realistic measurement noise. Unlike the Wigner function, which can exhibit negative values, the Husimi function is a smoothed version obtained via Gaussian convolution, making it an ideal tool for studying phase-sensitive quantum states, including squeezed and non-Gaussian states ~\cite{husimi1940some, takahashi1989distribution, appleby2000husimi}. Additionally, the $P$-distribution has been proposed, but it is known for having singular values, making it challenging for practical phase representation \cite{glauber1963coherent}. Furthermore, the Husimi phase probability distribution allows for intuitive visualization and efficient computation in polar coordinates, making it highly relevant for quantum metrology applications ~\cite{buvzek1995sampling}.

The study of phase distribution plays a crucial role in understanding the quantum properties of optical states \cite{chizhov1993phase,banerjee2007phase, banerjee2007phaseGR}. While phase distribution based on the phase operator formalism has been extensively explored for Gaussian and non-Gaussian states at higher average photon numbers ~\cite{malpani2020impact, malpani2019quantum,linowski2023formal}, our focus in this work is on the Husimi phase probability distribution as a means to understand the nonclassical and non-Gaussian quantum states. The Husimi phase distribution enables the identification of quantum features such as phase squeezing, quantum interference, and non-Gaussianity, which are crucial for quantum-enhanced technologies. We explore state engineering techniques, including photon addition, photon subtraction and photon catalysis, to manipulate the phase properties of single mode vacuum (SSV) states and two-mode squeezed vacuum (TMSV) states and examine whether the non-Gaussian (NG) operation localizes or delocalizes the Husimi phase distribution. Additionally, we investigate the impact of the amplitude damping channel on the Husimi phase distribution, providing insights into decoherence and quantum noise~\cite{banerjee2018open} effects on engineered quantum states. 


 The rest of the paper is structured as follows. In Sec.~\ref{sec:Preliminaries}, we motivate the phase space approach, based on the Husimi distribution function,  used subsequently in this work.  The Husimi function of the non-Gaussian SSV state (NGSSV) is derived in Sec.~\ref{sec:ngssv} and its phase distribution studied. Similarly, we focus our attention on the TMSV state in Sec.~\ref{sec:ngtmsv}, where we derive the Husimi function of the non-Gaussian TMSV state (NGTMSV) and study the corresponding phase distribution. In Sec. \ref{sec:NGSVS}, the dynamical evolution of the single mode and two mode squeezed vaccum states in an amplitude dampling channel is analyzed.  We briefly summarize the main results and discuss future prospects in Sec. \ref{sec: Results_Discussion}, and follow it up with our conclusions in Sec. \ref{conc}.

\section{Preliminaries}\label{sec:Preliminaries}
The $s$-parametrized phase distribution generalizes phase representation by connecting different quasi-probability distributions and is defined using an integral over the $s$-parametrized quasiprobability function \cite{lee1995theory}. 
The \(s\)-parametrized quasiprobability distribution can  be directly expressed using the density matrix \(\hat{\rho}\) as:
\begin{equation}
    W(\alpha, s) = \frac{1}{\pi^2} \int d^2\lambda \, \text{Tr}[\hat{\rho} D(\lambda)] e^{\alpha \lambda^* - \alpha^* \lambda} e^{s |\lambda|^2 /2},
\end{equation}
where \(D(\lambda) = e^{\lambda \hat{a}^\dagger - \lambda^* \hat{a}}\) is the displacement operator.
The parameter determines the smoothing level, where $s=1$ corresponds to the Glauber $P$-distribution function ($P$-function), $s=0$ corresponds to the Wigner function ($W$-function), $s=-1$ corresponds to the Husimi function ($Q$-function), and $s<-1$ represents additional noise effects that further smooth the phase distribution. The Measured phase distributions are well approximated by $s$-parametrized distributions with $s\leq -1$, and the Husimi function phase distribution provides a practical approximation for experimental data.

The three distributions ($P$, $W$ and $Q$) are related by a convolutional product, which is a modified Weierstrass transform. For instance, the Husimi distribution is the convolution product $\circledast$ of the Wigner distribution (state $\hat\rho$) and the Gaussian state centered at the origin of phase
 space (vacuum state) $W_{|0\rangle}$ ~\cite{fabre2020quantum,appleby2000husimi}:
\begin{equation}
    Q_{\hat{\rho}}(x,p)=W_{\hat{\rho}}\circledast W_{|0\rangle}= \int dq_1 dp_1 W_\text{NG}(q_1,p_1)     W_{|\alpha\rangle}(q_1,p_1). \label{eq:w2Q}
\end{equation}
The Husimi distribution, also called the Berezin function, is a smeared version of the Wigner
distribution, so that it becomes a positive distribution.

From both theoretical and experimental perspectives, phase distributions offer valuable insights into quantum optical properties. The phase distribution (\(P_Q\)) derived from the Husimi function, particularly in polar representation, provides an effective link between theoretical concepts and practical measurements \cite{leonhardt1995canonical}: 
\begin{equation}  
    P_Q(\theta) = \int_{0}^{\infty} da \, a \, Q(a\cos\theta, a\sin\theta). \label{eq:PQ}  
\end{equation}  

This formulation inherently accounts for the unavoidable effects of quantum measurement noise while preserving essential phase information. As a result, it plays a vital role in phase-sensitive quantum technologies and state estimation protocols and will be made use of in this work.

\section{Non-Gaussian single mode squeezed vacuum state}\label{sec:ngssv}
 In this section, we study the effect of different non-Gaussian operations on the Husimi phase distribution of the SSV state.
 We represent the mode of the SSV state by the quadrature operators $\hat{q}$ and $\hat{p}$ with the annihilation operations related to the quadrature operators via the relation $\hat{a}=(\hat{q}+i \hat{p})/\sqrt{2} $.
 The  SSV state is given by \cite{loudon2000quantum} 
\begin{equation}
    |\psi\rangle_\text{SSV} = \exp[r(\hat{a}^2-\hat{a}{^{\dagger}}^2)/2]|0\rangle,
\end{equation}
with $r$ being the squeezing parameter, which is in general a complex number.

\begin{figure}[!ht]
	
    \includegraphics[width=1\linewidth]{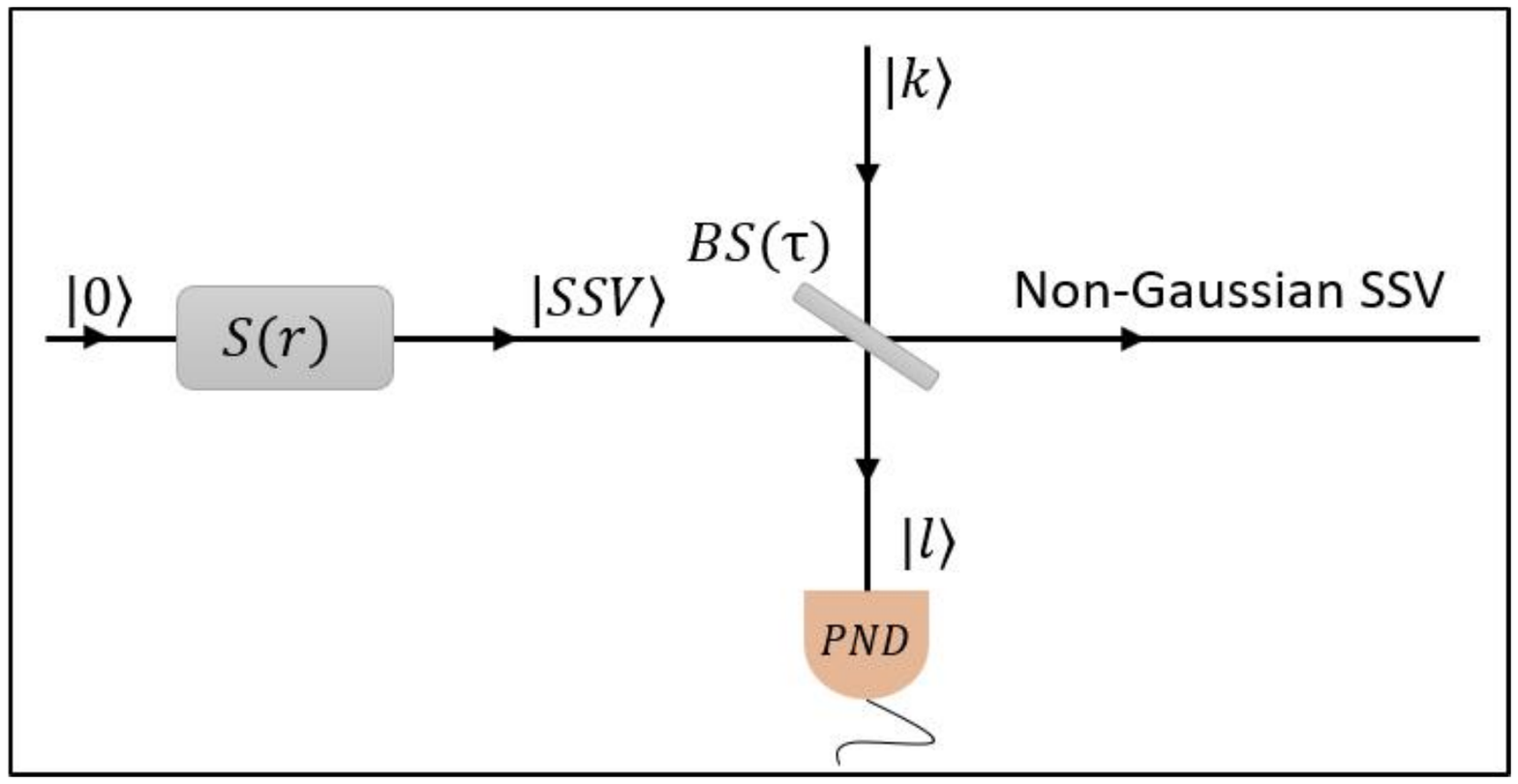}
	\caption{Schematic representation of a non-Gaussian operation on a single-mode squeezed vacuum (SSV) state. Where \( S(r) \): squeezing operator  generates the SSV state with squeezing parameter \( r \),  \( BS(\tau) \): beam splitter  with transmittance \( \tau \), and a PND: photon-number resolving detector.}
	\label{ngssv}
\end{figure}

To implement non-Gaussian operations on a single-mode squeezed vacuum (SSV) state, we combine the mode corresponding to the SSV state with an auxiliary mode initialized in the Fock state \(|k\rangle\). These two modes are then mixed using a beam splitter with transmissivity \(\tau\), as depicted in Fig.~\ref{ngssv}. Afterward, photon detection is performed on the output auxiliary mode. A successful detection of \(l\) photons signifies the execution of a non-Gaussian operation on the SSV state.  

Depending on the relation between \(k\) and \(l\), different operations are implemented on the input SSV state:  
  \begin{itemize}
      \item  If \(k < l\), a photon subtraction (PS) operation is performed.  
    \item  If \(k > l\), a photon addition (PA) operation is performed.  
    \item  If \(k = l\), a photon catalysis (PC) operation is performed.
  \end{itemize}

The resulting non-Gaussian state is given by  
\[
|\psi\rangle_{\text{NG-SSV}} \propto (\hat{a}^{\dagger k} \hat{a}^l) |\text{SSV}\rangle,
\]  
where \(\hat{a}^\dagger\) and \(\hat{a}\) represent the photon creation and annihilation operators, respectively, and \(k\) and \(l\) denote the number of photon additions and subtractions applied to the initial SSV state.

The  $Q$-distribution can be derived from  the convolution of the Wigner function and the vacuum state using Eq. \eqref{eq:w2Q}, as follow ~\cite{appleby2000husimi}:
	 \begin{equation}
	Q(q,p) =  \int dq_1 dp_1 W_\text{NG}(q_1,p_1)     W_{|\alpha\rangle}(q_1,p_1),
	 \end{equation}
where  $W_\text{NG}(q_1,p_1) $ is the Wigner function of the non-Gaussian SSV state and $W_{|\alpha\rangle}(q_1,p_1)$ is the Wigner function of the coherent state given by
\begin{equation}
    W_{|\alpha\rangle}(q_1,p_1) = \frac{1}{\pi}     \exp \left[ -(q-q_1)^2  -(p-p_1)^2\right].
\end{equation}

Using the Wigner function of the non-Gaussian SSV state~\cite{chandan-pra-23}, the Husimi function can be calculated as
  \begin{equation}
	Q(\xi) = \frac{\sqrt{1-\lambda ^2 \tau ^2}}{2 \pi } \dfrac{ \bm{\widehat{F}} \exp \left( \xi^T A_1 \xi + u^T A_2 \xi+ u^T A_3 u \right)}{  \bm{\widehat{F}} \exp \left(   u^T A_4 u \right)}, \label{eq5}
\end{equation}
where $\lambda= \tanh (r)$ and the column vectors  $\xi = (q,p)^T$ and $u=(u_1,v_1,u_2,v_2)^T$ with the differential operator $\bm{\widehat{F}} $ being 
\begin{equation}\label{Foperator}
	\bm{\widehat{F}} =   \frac{\partial^{k}}{\partial\,u_1^{k}} \frac{\partial^{k}}{\partial\,v_1^{k}} \frac{\partial^{l}}{\partial\,u_2^{l}} \frac{\partial^{l}}{\partial\,v_2^{l}}\{ \bullet \}_{\substack{u_1= v_1=0\\ u_2= v_2=0}},\\
\end{equation}
The matrix $A_1$, $A_2$, $A_3$ and $A_4$  are 
 \begin{equation}
     A_1= \frac{-1}{2}\left(
\begin{array}{cc}
 \lambda  \tau +1 & 0 \\
 0 & 1-\lambda  \tau  \\
\end{array}
\right),
 \end{equation}

\begin{equation}
     A_2= \frac{\sqrt{1-\tau}}{2} \left(
\begin{array}{cc}
 1 & i \\
 -1 & i \\
 \lambda  \sqrt{\tau } & -i \lambda  \sqrt{\tau } \\
 -\lambda  \sqrt{\tau } & -i \lambda  \sqrt{\tau } \\
\end{array}
\right),
 \end{equation}
 \begin{equation}
     A_3 =\frac{1}{4}\left(
\begin{array}{cccc}
 0 & 0 & 0 & -\sqrt{\tau } \\
 0 & 0 & -\sqrt{\tau } & 0 \\
 0 & -\sqrt{\tau } & \lambda  (\tau -1) & 0 \\
 -\sqrt{\tau } & 0 & 0 & \lambda  (\tau -1) \\
\end{array}
\right),
 \end{equation}
and
\begin{widetext}

\begin{equation}\label{A4}
     A_4=\frac{1}{4 \left(\lambda ^2 \tau ^2-1\right)}  \left(
\begin{array}{cccc}
 -\lambda  (\tau -1) \tau  & 1-\tau  & \lambda  (\tau -1) \sqrt{\tau } & \sqrt{\tau } \left(1-\lambda ^2 \tau \right) \\
 1-\tau  & -\lambda  (\tau -1) \tau  & \sqrt{\tau } \left(1-\lambda ^2 \tau \right) & \lambda  (\tau -1) \sqrt{\tau } \\
 \lambda  (\tau -1) \sqrt{\tau } & \sqrt{\tau } \left(1-\lambda ^2 \tau \right) & \lambda -\lambda  \tau  & -\lambda ^2 (\tau -1) \tau  \\
 \sqrt{\tau } \left(1-\lambda ^2 \tau \right) & \lambda  (\tau -1) \sqrt{\tau } & -\lambda ^2 (\tau -1) \tau  & \lambda -\lambda  \tau  \\
\end{array}
\right).
     \end{equation}

\end{widetext}
 In the   unit transmissivity limit  (\(\tau \to 1\)), the   generalized Husimi distribution function  for non-Gaussian modified squeezed vacuum states is derived from Eq.~\eqref{eq5}, incorporating   photon subtraction, photon addition, and photon catalysis. 
For the   photon subtraction case, where   no photon addition occurs  (\( k = 0 \)), the Husimi distribution function corresponds to the   \( l \)-photon-subtracted squeezed vacuum (\( l \)-PSSSV) state, represented as \( |\psi\rangle_{\text{PS}} \propto \hat{a}^l |\text{SSV}\rangle \).
On the other hand, in the   photon addition case, where   no photon subtraction occurs  (\( l = 0 \)), the Husimi distribution function corresponds to the   \( k \)-photon-added squeezed vacuum (\( k \)-PASSV) state, given by \( |\psi\rangle_{\text{PA}} \propto \hat{a}^{\dagger k} |\text{SSV}\rangle \). 

In the case of  photon catalysis, the squeezed vacuum state undergoes partial photon transfer ($k=l=m$), modifying its quantum coherence and phase properties. The $m$-photon-catalyzed squeezed vacuum (m-PCSSV) state  is expressed as \( |\psi\rangle_{\text{PC}} \propto (\hat{a}^{\dagger m} \hat{a}^m) |\text{SSV}\rangle \).

By setting \( k = 0 \), \( l = 0 \), and \( \tau = 1 \) in Eq.~(\ref{eq5}), we retrieve the   Husimi  function   corresponding to the   standard single-mode squeezed vacuum (SSV) state, which serves as the baseline case without any photon addition or subtraction, i.e.,

 \begin{equation}
 	Q (q, p) = \frac {\sqrt {1 - \lambda^2}} {2\pi}\exp\left[-q^2 (1 + \
 	\lambda)/2 - p^2 (1 - \lambda)/2 \right].
 \end{equation}
On integrating the Husimi function over the radial coordinate, we obtain Husimi phase distribution Eq. \eqref{eq:PQ}.

\begin{figure}[!ht]
	\includegraphics[scale=1]{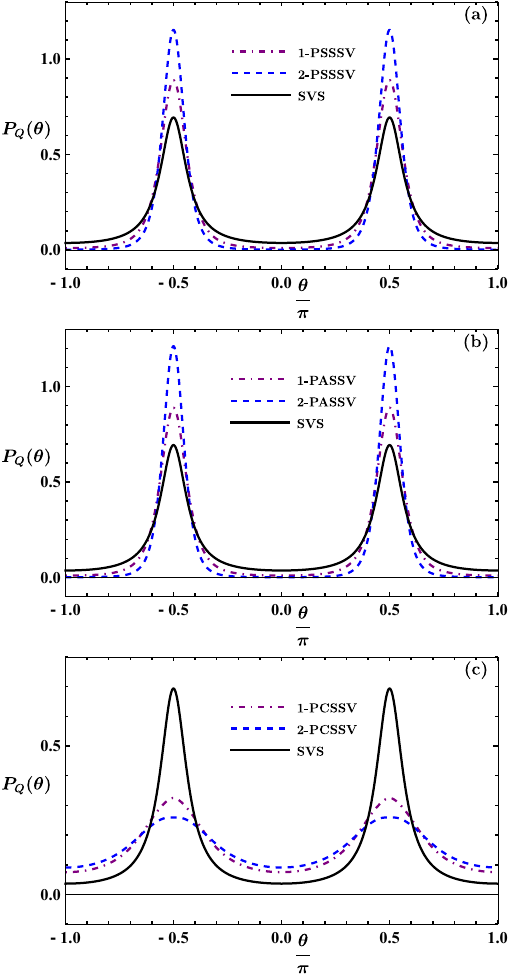}
	\caption{  Husimi phase distribution $P_Q(\theta)$ as a function of scaled phase parameter $\theta /\pi$ for different NGSSV states. We have set $\lambda=0.9$ and $\tau=0.9$.
	}
	\label{ngssv_phase}
\end{figure}

\begin{figure*}[!ht]
    \centering
    \begin{subfigure}[b]{0.32\textwidth}
        \includegraphics[width=\textwidth]{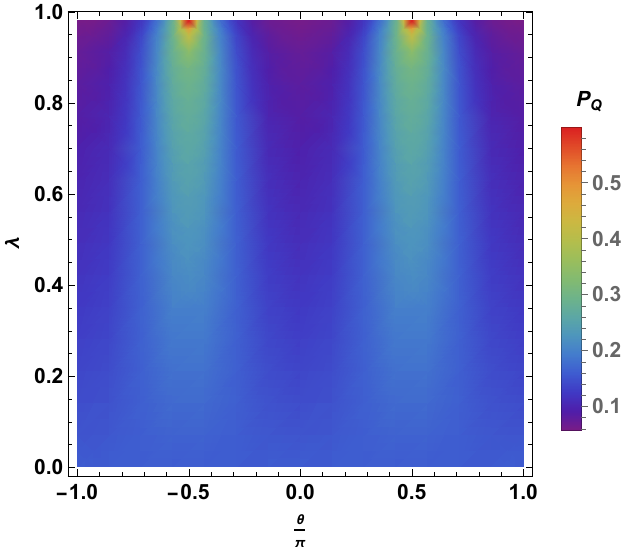}
        \caption{1-PC-SSV with $\lambda$ at $\tau=0.9$}
        \label{fig:sub4}
    \end{subfigure}
    \hfill
    \begin{subfigure}[b]{0.32\textwidth}
        \includegraphics[width=\textwidth]{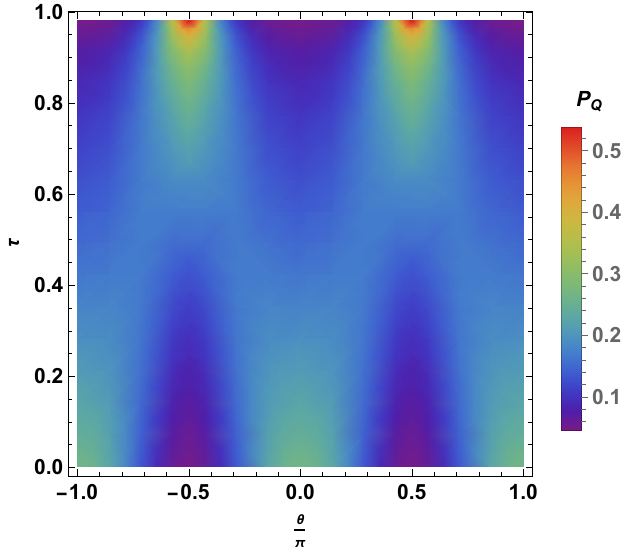}
        \caption{1-PC-SSV with $\tau$ at $\lambda=0.9$}
        \label{fig:sub5}
    \end{subfigure}
    \hfill
    \begin{subfigure}[b]{0.32\textwidth}
        \includegraphics[width=\textwidth]{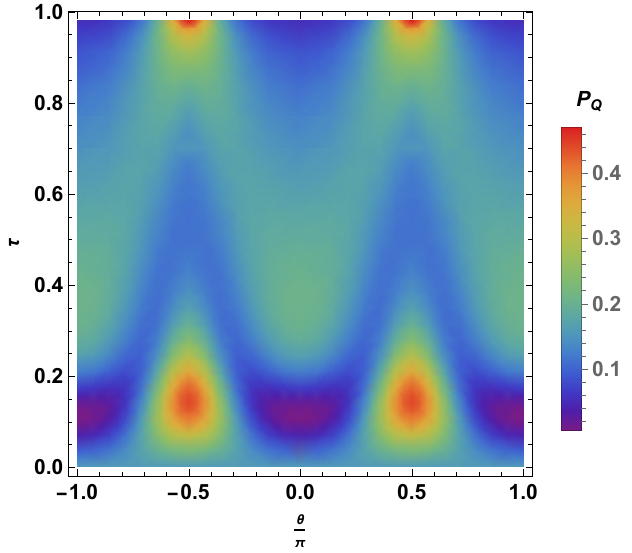}
        \caption{2-PC-SSV with $\tau$ at $\lambda=0.9$ }
        \label{fig:sub6}
    \end{subfigure}
    \begin{subfigure}[b]{0.32\textwidth}
        \includegraphics[width=\textwidth]{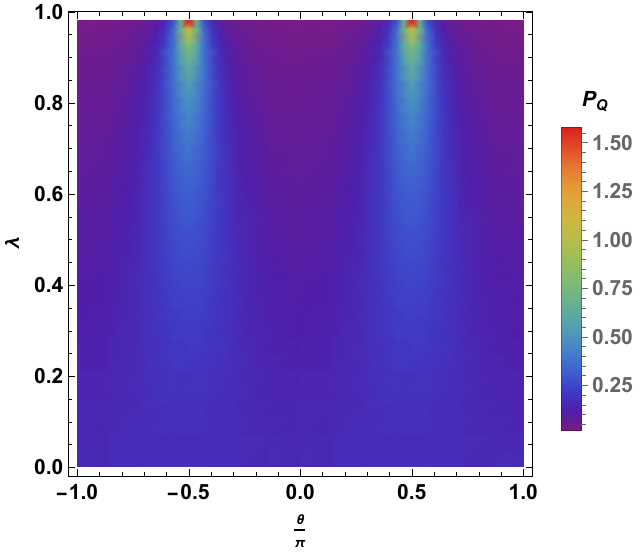}
        \caption{SSV with $\lambda$}
        \label{fig:sub1}
    \end{subfigure}
    \hfill
    \begin{subfigure}[b]{0.32\textwidth}
        \includegraphics[width=\textwidth]{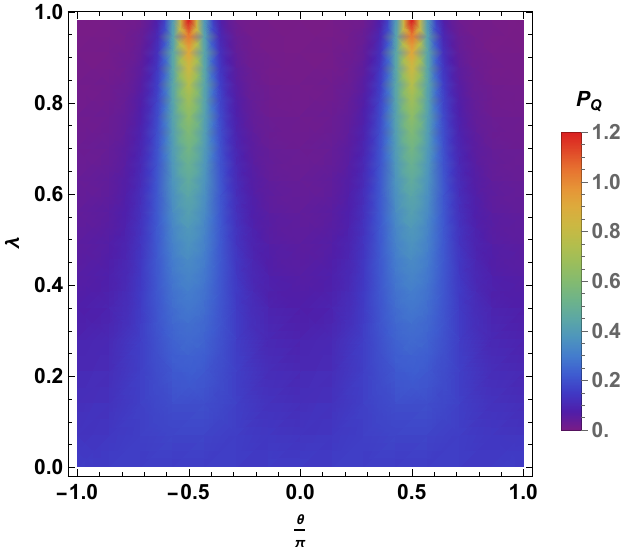}
        \caption{1-PS/PA-SSV with $\lambda$ at $\tau=0.9$}
        \label{fig:sub2}
    \end{subfigure}
    \hfill
    \begin{subfigure}[b]{0.32\textwidth}
        \includegraphics[width=\textwidth]{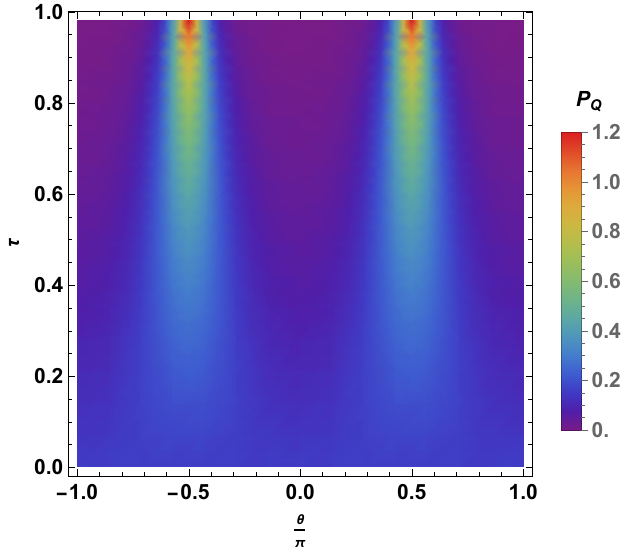}
        \caption{1-PS/PA-SSV with $\tau$ at $\lambda=0.9$ }
        \label{fig:sub3}
    \end{subfigure}

    \vspace{1cm} 

    \caption{Density plots of the Husimi phase distribution function for SSV states subjected to various non-Gaussian operations. The color bar indicates the intensity of the function. a) 1-PC applied to SSV: variation with squeezing parameter \( \lambda \) at fixed beam splitter transmittance \( \tau = 0.9 \), b) 1-PC-SSV: variation with \( \tau \) at fixed \( \lambda = 0.9 \), c) 2-PC applied to SSV: variation with \( \tau \) at fixed \( \lambda = 0.9 \), d) SSV without non-Gaussian operations: variation with \( \lambda \), (e) 1-PS/PA applied to SSV: variation with \( \lambda \) at fixed \( \tau = 0.9 \) and (f) 1-PS/PA-SSV: variation with \( \tau \) at fixed \( \lambda = 0.9 \).}
    \label{ngssv_density}
\end{figure*}

 For the SSV state, the Husimi phase distribution function is
\begin{equation}
  P_Q(\theta)_\text{SVS}=  \frac{\sqrt{1-\lambda ^2}}{2 \pi  (\lambda  \cos (2 \theta )+1)}.\label{Eq_0ssv}
\end{equation}
 At $\lambda=0$, i.e., for the case of zero squeezing,  the state is simply the vacuum state and the corresponding  Husimi phase distribution function, from 
 Eq.~\eqref{Eq_0ssv} is $P_Q(\theta)_{|0\rangle}=  1/(2 \pi)$.
\begin{equation}
P_Q(\theta)_\text{1-PS} =P_Q(\theta)_\text{1-PA} =	\frac{\left(1-\lambda ^2 \tau ^2\right)^{3/2}}{2 \pi  (\lambda  \tau  \cos (2 \theta )+1)^2}.\label{Eq_1ps/pc}
\end{equation}
In the limit $\tau\rightarrow1$, the 1-PSSSV state is 
  \begin{equation}
      \mathcal{N}_1 \hat{a} \exp[r(\hat{a}^2-\hat{a}{^{\dagger}}^2)/2] |0\rangle  = \mathcal{N}_2 \exp[r(\hat{a}^2-\hat{a}{^{\dagger}}^2)/2] |1\rangle,
  \end{equation}
  where $\mathcal{N}_i$'s are normalization factor. When $r=0$ $  (\lambda=0)$, the state is $|1\rangle$ and the corresponding  Husimi phase distribution function from 
Eq.~\eqref{Eq_1ps/pc} is $P_Q(\theta)_{|1\rangle}=  1/(2 \pi)$.
We have shown the  Husimi phase distribution for different non-Gaussian states in Fig.~\ref{ngssv_phase}. The results show that while PS and PA operations localizes the Husimi phase distribution of the SSV state, PC operations delocalizes. As we subtract or add more photons, the Husimi phase distribution gets more localized.\\

The phase distribution \( P_Q(\theta) \) of a photon-catalyzed squeezed vacuum state exhibits distinct oscillatory structures that depend on the phase parameter \( \theta \) and the beam splitter transmittance \( \tau \). However, with squeezing parameter \( \lambda \), it increases monotonically (see Fig. \ref{ngssv_density}a-c). In the case of single-photon catalysis, the distribution shows two prominent peaks around \( \theta/\pi = \pm 0.5 \), highlighting strong phase sensitivity due to quantum interference effects, with peak sharpness increasing as \( \tau \) grows. For two-photon catalysis, additional peaks emerge, creating a more intricate phase structure due to higher-order non-Gaussian features. Additionally, increasing the squeezing parameter \( \lambda \) enhances phase sensitivity, leading to more pronounced and sharper peaks. Notably, in both photon-added and photon-subtracted cases (see Fig. \ref{ngssv_density}d-f), the phase distribution peaks become more prominent with increasing squeezing and beam splitter transmittance, emphasizing the role of these parameters in controlling quantum state phase properties.
 
\section{Non-Gaussian two mode squeezed vacuum state}\label{sec:ngtmsv}
We now turn to analyze the effect of non-Gaussian operation on the Husimi phase distribution of TMSV state.
We represent the two modes of the TMSV state via the quadrature operators $\hat{q}_1$, $\hat{p}_1$, $\hat{q}_2$, and $\hat{p}_2$. 
The TMSV state is given by
  \begin{equation}
  |\text{TMSV} \rangle = \exp[r(\hat{a}_1^\dagger \hat{a}_2^\dagger-  \hat{a}_1  \hat{a}_2 )]|0\rangle |0\rangle,
  \end{equation}
  where $r$ is the squeezing parameter (two mode).

\begin{figure}[!ht]
  \includegraphics[scale=1]{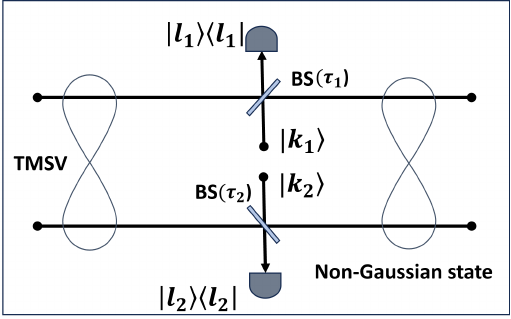}
  \caption{  Schematic representation of the generation of a non-Gaussian state from a two-mode squeezed vacuum (TMSV) state.  Where \( BS(\tau_1) \) and \( BS(\tau_2) \): beam splitters  with transmittance \( \tau_1 \) and \( \tau_1 \)}
  \label{ngtmsv}
\end{figure}

 In order to perform non-Gaussian operation on one mode, say mode $A_1$, we mix it with  an auxiliary mode initialized to Fock state $|k_1\rangle$ using a beam splitter of transmissivity $\tau_1$. Subsequently, photon detection is performed on the output auxiliary mode.  A successful detection of $l_1$ photons heralds the implementation of non-Gaussian operations. In  a similar way, non-Gaussian operation can be executed on the other mode. The schematic is shown in Fig.~\ref{ngtmsv}. For $k_1<l_1$, $k_1>l_1$ and $k_1=l_1$, PS, PA, and PC operations are implemented on the mode $A_1$ of the TMSV state (similarly on $A_2$ mode). 
 
A generalized non-Gaussian (NG) state is obtained by applying photon addition (\(\hat{a}^\dagger\)), photon subtraction (\(\hat{a}\)), and photon catalysis (\(\hat{a}^{\dagger k} \hat{a}^k\)) to each mode of the TMSV state. This can be expressed as \[
|\psi\rangle_{\text{NG-TMSV}} \propto (\hat{a}_1^{\dagger k_1} \hat{a}_1^{l_1}) (\hat{a}_2^{\dagger k_2} \hat{a}_2^{l_2}) |\text{TMSV}\rangle.
\]  

In the symmetric (Sym) case, identical operations are applied to both modes, leading to the conditions  $k_1 = k_2 = k$  and $l_1 = l_2 = l$. Conversely, in the asymmetric (Asym) case, the number of photon additions and subtractions differs between the modes, meaning $k_1 \neq k_2$ and $ l_1 \neq l_2$. The special cases for photon operations on TMSV are provided in Table \ref{table1}.

The Husimi function for NGTMSV state can be calculated in Wigner distribution formalism as follows (using Eq. \eqref{eq:w2Q}):
  \begin{equation}
  	Q(\xi_1,\xi_2) =  \int d^2 \xi_3 d^2 \xi_4 W_\text{NG-TMSV}(\xi_3,\xi_4)     W_{|\alpha\rangle}(  \xi_3) W_{|\alpha\rangle}(  \xi_4).
  \end{equation}
 where $\xi_i =(q_i,p_i)^T$.

Using the Wigner function ~\cite{crs-ngtmsv-met}, we can calculate the Husimi distribution as
  \begin{equation}\label{eq_tmsv_Q_phase}
	Q(\xi_1,\xi_2)=  \frac{1-\lambda ^2  \tau_1 \tau_2}{4 \pi ^2} \dfrac{\bm{\widehat{F}_1} \exp \left(\bm{\xi}^T M_1 \bm{\xi}+\bm{u}^T M_2 \bm{\xi} + \bm{u}^T M_3 \bm{u} \right)}{\bm{\widehat{F}_1} \exp \left(\bm{u}^T M_4 \bm{u}\right)},
\end{equation}
with $\bm{u}=(u_1,v_1,u_2,v_2,u_1',v_1',u_2',v_2')^T$  
  and   $\bm{\widehat{F}_1} $ is defined as 
\begin{multline}
	\bm{\widehat{F}_1} =   \frac{\partial^{k_1}}{\partial\,u_1^{k_1}} \frac{\partial^{k_1}}{\partial\,v_1^{k_1}} \frac{\partial^{k_2}}{\partial\,u_2^{k_2}} \frac{\partial^{k_2}}{\partial\,v_2^{k_2}}\\
	\times \frac{\partial^{l_1}}{\partial\,u_1'^{l_1}} \frac{\partial^{l_1}}{\partial\,v_1'^{l_1}} \frac{\partial^{l_2}}{\partial\,u_2'^{l_2}} \frac{\partial^{l_2}}{\partial\,v_2'^{l_2}} \{ \bullet \}_{\substack{u_1= v_1=u_2= v_2=0\\ u_1'= v_1'=u_2'= v_2'=0}}.\\
\end{multline}
The matrix $M_1$ is 
\begin{equation}
    M_1=\frac{-1}{2}\left(
\begin{array}{cccc}
 1 & 0 & -\lambda  \sqrt{\tau_1} \sqrt{\tau_2} & 0 \\
 0 & 1 & 0 & \lambda  \sqrt{\tau_1} \sqrt{\tau_2} \\
 -\lambda  \sqrt{\tau_1} \sqrt{\tau_2} & 0 & 1 & 0 \\
 0 & \lambda  \sqrt{\tau_1} \sqrt{\tau_2} & 0 & 1 \\
\end{array}
\right).
\end{equation}

Further,   the matrices $M_2$, $M_3$, and $M_4$ are provided in Eqs.~(\ref{mat2}), (\ref{mat3}), and~(\ref{mat4}) of Appendix~\ref{appwigner}. We now bring out special cases of the general case derived in Eq.~(\ref{eq_tmsv_Q_phase}). 
 \begin{table}[!ht]
	\centering
	\caption{\label{table1}
		Special cases of ideal PS and PA operations}
	\renewcommand{\arraystretch}{1.5}
	\begin{tabular}{ |c |c |c|}
		\hline \hline
		Parameters value &State  & Operation\\
		\hline \hline
		$k_1=k_2=l_2=0$, $\tau_2\rightarrow1$ &   $  \propto \hat{a}_1^{l_1} |\text{TMSV}\rangle$ & Asym $l_1$-PS  \\ \hline
		$k_2=l_1=l_2=0$, $\tau_2\rightarrow1$ &  $  \propto \hat{a}_1{^\dagger}^{k_1} |\text{TMSV}\rangle$  & Asym $k_1$-PA \\  \hline
		$k_1=k_2=0$, $l_1=l_2=l$  &   $  \propto \hat{a}_1^{l} \hat{a}_2^{l} |\text{TMSV}\rangle$  & Sym $l$-PS\\  \hline
		$l_1=l_2=0$, $k_1=k_2=k$  &     $  \propto \hat{a}_1{^\dagger}^{k} \hat{a}_2{^\dagger}^{k} |\text{TMSV}\rangle$ & Sym $k$-PA \\  \hline \hline
	\end{tabular}
\end{table}

   Setting $\tau_1=\tau_2=1$ with $k_1=k_2=l_1=l_2=0$ in Eq.~\eqref{eq_tmsv_Q_phase} yields the Husimi function of the TMSV state:
    \begin{equation}
    	\begin{aligned}
    		Q(\xi_1,\xi_2)= \frac{1-\lambda^2}{4 \pi^2}	&\exp  \bigg[ -\big(q_1^2 + q_2^2 - 2\lambda q_1 q_2 \\
    		&+ p_1^2 + 
    		p_2^2 - 2\lambda p_1 p_2 \big)/2  \bigg].
    	\end{aligned}
    \end{equation}

   \begin{figure}[!ht]
 	\includegraphics[scale=1]{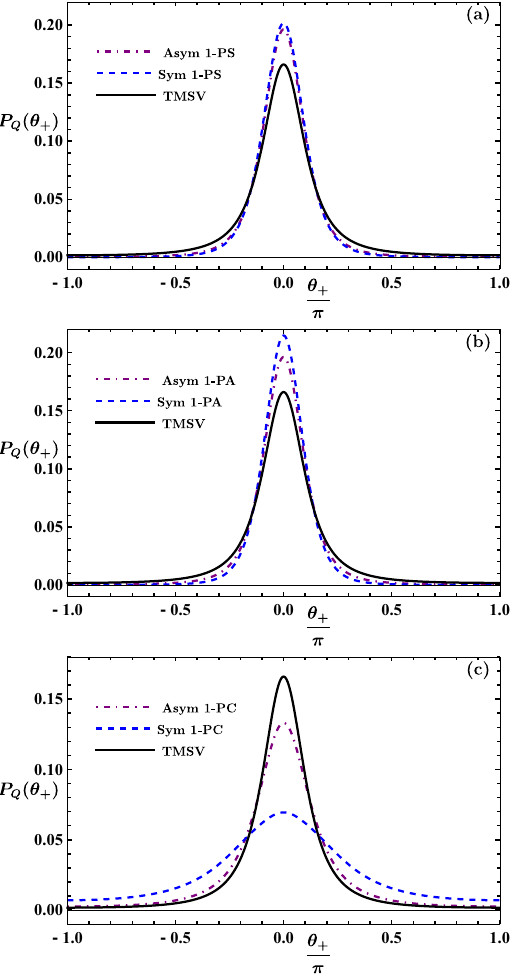}
 	\caption{   Husimi phase distribution $P_Q(\theta_+ = \theta_1+\theta_2)$ as a function of scaled phase parameter $\theta_+ /\pi$ for different NGTMSV  states.  We have set $\lambda=0.9$ and $\tau=0.9$.
 	}
 	\label{ngtmsv_phase}
 \end{figure}
 
 On integrating the Husimi distribution over the radial coordinates, we obtain Husimi phase distribution defined as
\begin{equation}
	\begin{aligned}
	 P_Q(\theta_1,\theta_2) = \int_{0}^{\infty} \int_{0}^{\infty} da_1\;da_2\; a_1a_2\,& Q(a_1 \cos \theta_1, a_1 \sin \theta_1,\\
	 &a_2 \cos \theta_2, a_2 \sin \theta_2).
	\end{aligned}
     \end{equation}
 The phase distribution function for the TMSV state turns out to be
  \begin{equation}
 P(\theta_1,\theta_2)=\frac{\left(1-\lambda ^2 \right) \left(C_1 \lambda  \left(2 \tan ^{-1}\left(\frac{C_1 \lambda }{\sqrt{C_0}}\right)+\pi \right)+2 \sqrt{C_0}\right)}{8 \pi ^2 C_0^{3/2}},
  \end{equation}
 $C_0=1-\lambda ^2 \cos ^2(\theta_1+\theta_2)$ and $C_1 =\cos (\theta_1+\theta_2) $. The phase distribution function for different non-Gaussian TMSV states including the TMSV state is a function of $\theta_+ = \theta_1+\theta_2$. The phase distribition function  \( P_Q(\theta_+) \) for different non-Gaussian TMSV states as a function of $\theta_+$ are depicted in Fig.~\ref{ngtmsv_phase}. These highlight the effects of photon subtraction (PS), photon addition (PA), and photon catalysis (PC) on phase correlations. The TMSV state (black solid line) serves as a reference, displaying a peak at \( \theta_+ = 0 \), which signifies strong  phase synchronization  between the two modes due to entanglement. Notably, for all cases, the peak remains centered at \( \theta_+ = 0 \), indicating that the fundamental phase correlations in TMSV states persist even after non-Gaussian modifications.
 The effect of non-Gaussian operations on the TMSV is seen to follow a similar trend as that of non-Gaussian operation on SSV state (Figs.~\ref{ngssv_phase}  $\&$ ~\ref{ngssv_density}).

The application of non-Gaussian operations significantly alters the  sharpness and intensity  of the phase distribution, with  symmetric operations (blue dashed curves) consistently showing a more pronounced peak  compared to asymmetric cases (dot-dashed pink curves). This suggests that applying the same operation to both modes enhances  quadrature correlations, reinforcing phase synchronization. In photon subtraction (Fig. \ref{ngtmsv_phase}a), the Husimi distribution exhibits a  narrower peak, as compared to TMSV case, indicating an enhancement in phase correlations, as photon subtraction is known to increase nonclassicality. Also, photon addition (Fig. \ref{ngtmsv_phase}b) leads to a sharper peak.
Photon catalysis, demonstrates a distinct behavior. Even though, it can be seen that the overall shape of the phase distribution is delocalized, it can be shown that the  peak position remains unchanged  regardless of beam splitter transmittance \( \tau \) (density plot not attached here), in contrast to the case of photon catalysis of SSV states (see Fig. \ref{ngssv_density}b-c).

\section{Dynamical Evolution of Non-Gaussian Squeezed Vacuum State (NGSVS) in Amplitude Damping Channel}\label{sec:NGSVS}
The dynamics of an open quantum system interacting with its environment can be described using a Lindblad master equation. For a multimode system with $N$ non-interacting modes experiencing amplitude damping noise \cite{chen2006simultaneous,srikanth2008squeezed, omkar2013dissipative}, the master equation can be written as 

\begin{align}
   \frac{\partial \rho(t)}{\partial t}=\sum_{i=1}^{N} \gamma_i \left[\hat{a}_i\rho(t)\hat{a}_i^{\dag}-\frac{1}{2}\left(\hat{a}_i^{\dag}\hat{a}_i\rho(t)+\rho(t)\hat{a}_i^{\dag}\hat{a}_i\right)\right] \label{eq:multimaster}.
\end{align}
Here, $\rho(t)$ is the density matrix of the system at time $t$, $\hat{a}_i$ and $\hat{a}_i^{\dagger}$ are the annihilation and creation operators for mode $i$, and $\gamma_i$ is the damping rate for mode $i$.
The Lindblad operator for $i^{th}$-mode  is $\hat{a}_i$. 
Using this, the master Eq. \eqref{eq:multimaster} can be rewritten as
\begin{align}
    \frac{\partial \rho(t)}{\partial t} = \sum_{i=1}^{N} \frac{\gamma_i}{2} \mathcal{L}_i \rho(t),
\end{align}
where, the action of superoperator $\mathcal{L}$ can be defined as
\begin{equation}
 \mathcal{L}_{i}\rho(t)= \left[2\hat{a}_i\rho(t)\hat{a}_i^{\dag}-\left(\hat{a}_i^{\dag}\hat{a}_i\rho(t)+\rho(t)\hat{a}_i^{\dag}\hat{a}_i\right)\right].
\end{equation}

Assuming that the super-operators $\mathcal{L}_i$ for different modes commute, i.e., $[\mathcal{L}_i, \mathcal{L}_j] = 0$ for $i \neq j$, the solution can be expressed as:
\begin{align}
    \rho(t) = \exp\left(\sum_{i=1}^{N} \frac{\gamma_i t}{2} \mathcal{L}_i\right)\rho(0).
\end{align}
This can be further factorized due to the commutation property:
\begin{align}
    \rho(t) = \left(\prod_{i=1}^{N} e^{\frac{\gamma_i t}{2} \mathcal{L}_i}\right) \rho(0) \label{eq:multimaster_sol}.
\end{align}
The action of the super-operator $e^{\frac{\gamma_i t}{2} \mathcal{L}_i}$ on $\rho(0)$ for mode $i$ can be explicitly written as:
\begin{align}
    e^{\frac{\gamma_i t}{2} \mathcal{L}_{i}} \rho(0)=\sum_{n=0}^{\infty} \left( \frac{[e^{\gamma_i t} - 1]^{n}}{n!} \hat{a}_i^{n} e^{-\frac{\gamma_i t}{2}  \hat{a}^\dagger \hat{a}_i} \rho(0) e^{-\frac{\gamma_i t}{2}   \hat{a}_i^\dagger \hat{a}_i} \hat{a}_i^{\dagger n} \right) \label{eq:multimaster_sol_term}.
\end{align}

The amplitude-damping channel describes energy dissipation, such as photon loss in an optical system, in particular optical fiber channel~\cite{chen2006simultaneous}. The commutativity of the super operators $\mathcal{L}_i$ implies that the modes evolve independently, allowing us to treat each mode separately.

\subsection{NGSVS IN
AMPLITUDE DAMPING CHANNEL}
For the single-mode case, the solution simplifies to (using Eq. \eqref{eq:multimaster_sol} and \eqref{eq:multimaster_sol_term}):
\begin{align}
     \rho(t)=\sum_{n=0}^{\infty} \left( \frac{[e^{\gamma_1 t} - 1]^{n}}{n!} \hat{a}_1^{n} e^{-\frac{\gamma_1 t}{2}  \hat{a}_1^\dagger \hat{a}_1} \rho(0) e^{-\frac{\gamma_1 t}{2}   \hat{a}_1^\dagger \hat{a}_1} \hat{a}_1^{\dagger n} \right).
\end{align}

Thus, for an arbitrary initial element \( |\alpha\rangle\langle\alpha| \), its time evolution becomes:
\begin{equation}
    |\alpha\rangle\langle\alpha|\longrightarrow (|\alpha\rangle\langle\alpha|)_{t}= |\alpha \sqrt{\eta(t)}\rangle\langle\alpha \sqrt{\eta(t)}|, \label{coherentevolution}
\end{equation}

where $\eta(t)=e^{-\gamma_1 t}$
(derivation can be found in Appendix B).

In the coherent state basis, the NGSSV state (at $t=0$) can be written as  ~\cite{glauber1963coherent}:
\begin{equation}
    \rho_{NGSSV}=\int{{d^2\alpha }P(\alpha)|\alpha\rangle\langle\alpha|} \label{eq:NGSSV_sol}.
\end{equation}
Using Eqs. \eqref{coherentevolution} and \eqref{eq:NGSSV_sol}, the NGSSV state at time \(t\) is given by:
\begin{align}
&\rho_{NGSSV}(t)=\int{\frac{d^2\alpha}{\pi}P(\alpha)
|\alpha \sqrt{\eta(t)}\rangle\langle\alpha \sqrt{\eta(t)}|}.
\end{align}
The time-dependent Husimi \(Q\)-function can then be expressed as (derivation can be found in Appendix C)
\cite{Gerry_Knight_2004_chap3}:
\begin{figure}[!ht]
 	\includegraphics[width=\linewidth]{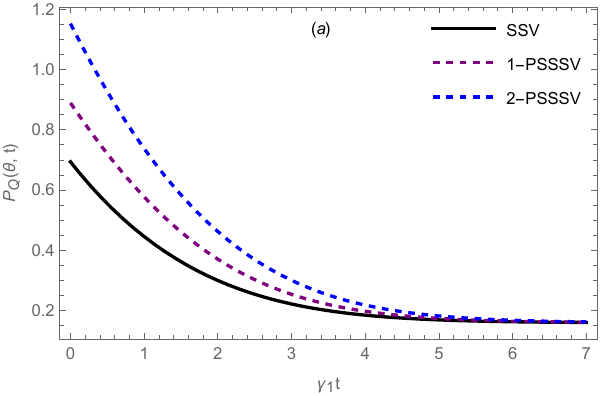}
    \includegraphics[width=\linewidth]{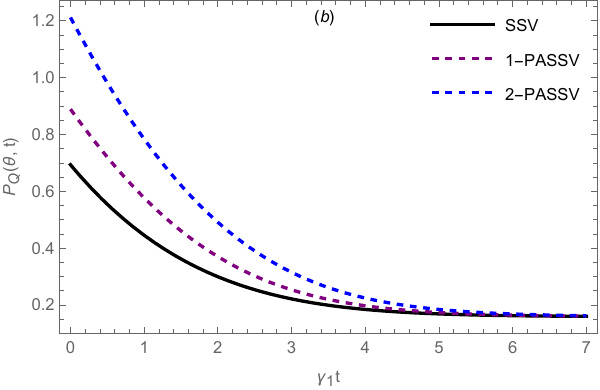}
    \includegraphics[width=\linewidth]{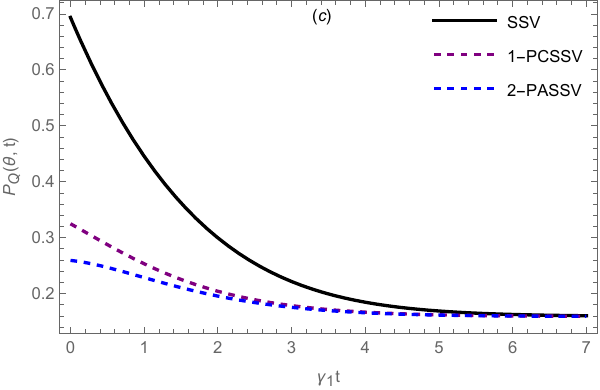}
 	\caption{   Husimi phase distribution $P_Q(\theta,t)$ as a function of scaled phase parameter $t$ for different NGSSV  states.  We have set $\lambda=0.9$ , $\tau=0.9$ and $\theta=\pm \pi/2$.
 	}
 	\label{ngssv_phase_dynamics}
 \end{figure}
\begin{align}
    Q(\xi,t)&=\frac{\langle \xi|\rho_{NGSSV}|\xi\rangle}{\pi} \nonumber\\
   &  = \int \frac{d^2 \beta}{\pi \left( 1 -\eta(t) \right)} \exp\left[ -\frac{|\xi - \beta \sqrt{\eta(t)}|^2}{1 - \eta(t)} \right] Q(\beta). 
\end{align}

By integrating the Husimi function over the radial coordinate,  we obtain the Husimi phase distribution:
\begin{equation}
    P_{Q}(\theta,t)=\int_{0}^{\infty} {da\: a\: Q(a\cos{(\theta)},a \sin{(\theta)},t)}.
\end{equation}
For the SSV state, the Husimi phase distribution dynamics can be represented by 
\begin{equation}
    P_{Q}(\theta,t)_{SSV}=\frac{\sqrt{(1 - \lambda^2) \big((1 + \eta(t))^2 - (1 - \eta(t))^2 \lambda^2\big)}}
    {2 \pi 
    \big(1 + \eta(t) - (1 - \eta(t)) \lambda^2 + 2  \eta(t) \lambda \cos(2\theta)\big)}. \label{Eq:PQ_ssv}
\end{equation}

Similarly, for the 1-PSSSV and 1-PASSV state, the Husimi phase distribution dynamics can be represented by
\begin{widetext}
\begin{equation}
    P_{Q}(\theta,t)_{1-PS}=P_{Q}(\theta,t)_{1-PA}=\frac{
    \sqrt{\frac{1 + \eta(t)}{1 - \eta(t)} + \lambda \tau} \, (1 - \lambda^2 \tau^2)^{3/2} 
    \big((1 + \eta(t))^3 - (1 - \eta(t))^3 \lambda^2 \tau^2 - 
    2 \eta(t) (1 - \eta(t)^2) \lambda \tau \cos(2\theta)\big)
    }{
    2 \pi 
    \sqrt{\frac{1 + \eta(t)}{1 - \eta(t)} - \lambda \tau} 
    \big(1 + \eta(t) + (1 - \eta(t)) \lambda \tau\big) 
    \big(1 + \eta(t) - (1 - \eta(t)) \lambda^2 \tau^2 + 
    2 \eta(t) \lambda \tau \cos(2\theta)\big)^2
    }.\label{Eq:PQ_1ps/pa_ssv}
\end{equation}

Equations~\eqref{Eq:PQ_ssv} and~\eqref{Eq:PQ_1ps/pa_ssv} reduce to Eqs.~\eqref{Eq_0ssv} and~\eqref{Eq_1ps/pc}, respectively, when the loss terms are set to zero ($\gamma_1=0$), i.e.,  by setting 
 $\eta(t)=1$. This is a nice consistency check for the calculations.

\end{widetext}

In the Fig. \ref{ngssv_phase_dynamics}, we have compared the behavior of the squeezed vacuum state (SSV)   with photon-subtracted (PSSSV), photon-added (PASSV), and photon-catalyzed (PCSSV) squeezed states under the influence of amplitude damping channel. It can be seen that the prominent peak value of phase distribution decreases monotonically as the scaled time \( \gamma_1 t \) increases, reflecting the effect of loss  due to amplitude damping. The NGSSV states (dashed curves) are seen in Fig. \ref{ngssv_phase_dynamics}a to exhibit a slower decay compared to the standard SSV, Eq.~(\ref{Eq:PQ_ssv}), indicating that non-Gaussian operations enhance phase robustness against decoherence. Notably, the 2-PSSSV ($P_Q(\theta)$ expression is cumbersome and hence not shown here)  maintains a higher phase distribution over time compared to the 1-PSSSV case (Eq.~\eqref{Eq:PQ_1ps/pa_ssv}), suggesting that increasing the number of photon operations further stabilizes phase properties.

Similar to the PSSSV state, the PASSV states follow the same trend, as can be observed from  Fig.~\ref{ngssv_phase_dynamics}b, where the 2-PASSV case shows stronger resilience than its single-photon counterpart. However, the PCSSV behaves differently, Fig.~\ref{ngssv_phase_dynamics}c, with an initially higher phase distribution but a slightly faster decay rate compared to photon-subtracted and photon-added cases. This suggests that while photon catalysis enhances initial phase sensitivity, it does not provide as much long-term robustness against amplitude damping.

\subsection{NGTMSV in an Amplitude Damping Channel }

For the two-mode case, the evolution of the quantum state under the amplitude damping channel can be expressed by using the action of super operator $e^{\frac{\gamma_i t}{2} \mathcal{L}_{i}} $ on each mode. The total state evolution is given by:
\begin{align}
 \rho(t) = \left(\prod_{i=1}^{2} e^{\frac{\gamma_i t}{2} \mathcal{L}_{i}}\right) \rho(0),
\end{align}

For the explicit form of the two-mode state evolution, we first apply the amplitude damping operation on the first mode, followed by the second mode. The resulting state is given by:
\begin{multline}
 \rho(t) = \sum_{n=0}^{\infty}  \frac{[e^{\gamma_2 t} - 1]^{n}}{n!} \hat{a}_2^{n} e^{-\frac{\gamma_2 t}{2} \hat{a}_2^\dagger \hat{a}_2} \\
 \times\left(e^{\frac{\gamma_1 t}{2} \mathcal{L}_{1}} \rho(0)\right) 
 e^{-\frac{\gamma_2 t}{2} \hat{a}_2^\dagger \hat{a}_2} \hat{a}_2^{\dagger n} ,
\end{multline}
where \(\gamma_1\) and \(\gamma_2\) are the damping rates for the first mode and second modes, respectively, and \(e^{\frac{\gamma_1 t}{2} \mathcal{L}_{1}} \rho(0)\) is the state evolved under the action of the damping channel on the first mode.

The explicit form of the first mode’s evolution is given by:
\begin{align}
 e^{\frac{\gamma_1 t}{2} \mathcal{L}_{1}} \rho(0) = \sum_{n=0}^{\infty}  \frac{[e^{\gamma_1 t} - 1]^{n}}{n!} \hat{a}_1^{n} e^{-\frac{\gamma_1 t}{2} \hat{a}_1^\dagger \hat{a}_1} \rho(0) 
 e^{-\frac{\gamma_1 t}{2} \hat{a}_1^\dagger \hat{a}_1} \hat{a}_1^{\dagger n}.
\end{align}

For an  initial two-mode coherent state \(|\alpha_1, \alpha_2\rangle\langle\alpha_1, \alpha_2|\), the evolution under the damping process can be obtained similar to the single-mode case. For mode 1, we have:
\[
    |\alpha_1\rangle\langle\alpha_1| \rightarrow |\alpha_1 e^{-\frac{\gamma_1 t}{2}}\rangle \langle \alpha_1 e^{-\frac{\gamma_1 t }{2}}|,
\]
and for mode 2:
\[
    |\alpha_2\rangle\langle\alpha_2| \rightarrow |\alpha_2 e^{-\frac{\gamma_2 t}{2}}\rangle \langle \alpha_2 e^{-\frac{\gamma_2 t}{2}}|.
\]
Therefore, the time evolution of the two-mode coherent state is:
\begin{multline} 
    |\alpha_1, \alpha_2\rangle\langle\alpha_1, \alpha_2| \rightarrow (|\alpha_1, \alpha_2\rangle\langle\alpha_1, \alpha_2|)_{t}\\
    = \left|\alpha_1 e^{-\frac{\gamma_2 t}{2}}, \alpha_2 e^{-\frac{\Gamma_2(t)}{2}}\left\rangle \right\langle \alpha_1 e^{-\frac{\gamma_1 t}{2}}, \alpha_2 e^{-\frac{\gamma_2 t}{2}}\right|.
\end{multline}
The initial $(t=0)$ two-mode squeezed vacuum state in the coherent-state representation can be written as:
\begin{equation}
    \rho_{\text{NGTMSV}} = \int {d^2 \alpha_1} {d^2 \alpha_2} P(\alpha_1, \alpha_2) |\alpha_1, \alpha_2\rangle \langle \alpha_1, \alpha_2|.
\end{equation}

Using the two-mode evolution equation derived above, the state at time \(t\) is:
\begin{multline}
    \rho_{\text{NGTMSV}}(t) = \int {d^2 \alpha_1} {d^2 \alpha_2} P(\alpha_1, \alpha_2)\\
    \times
    \left|\alpha_1 e^{-\frac{\gamma_1 t}{2}}, \alpha_2 e^{-\frac{\gamma_2 t}{2}}\left\rangle   
    \right\langle \alpha_1 e^{-\frac{\gamma_1 t}{2}}, \alpha_2 e^{-\frac{\gamma_2 t}{2}}\right|.
\end{multline}

The Husimi function for the two-mode squeezed state evolves similar to the single-mode case. The time-dependent two-mode Husimi function is:
\begin{multline}
 Q(\alpha_1,\alpha_2, t) = \int \frac{d^2 \beta_1 d^2 \beta_2\;Q(\beta_1,\beta_2)}{\pi^2 \left( 1 - e^{-\Gamma_1(t)} \right) \left( 1 - e^{-\Gamma_2(t)} \right)}\\
\quad \times\exp\left[ -\frac{|\alpha_1 - \beta_1 e^{-\frac{\Gamma_1(t)}{2}}|^2 }{\left(1 - e^{-\Gamma_1(t)}\right)} -\frac{|\alpha_2 - \beta_2 e^{-\frac{\Gamma_2(t)}{2}}|^2}{\left(1 - e^{-\Gamma_2(t)}\right)} \right] . \label{Qt_NGTMSV0} 
\end{multline}

Finally, by integrating the two-mode Husimi function over the radial coordinates, we obtain the Husimi phase distribution for the two-mode system:

\begin{multline}
    P_Q(\theta_1, \theta_2, t) = \int \int a_1 da_1 \,a_2 da_2 \,  \,\\
    \times \ Q(a_1 \cos(\theta_1), a_1 \sin(\theta_1), a_2 \cos(\theta_2), a_2 \sin(\theta_2), t).
\end{multline}

\begin{figure}
 	\includegraphics[width=\linewidth]{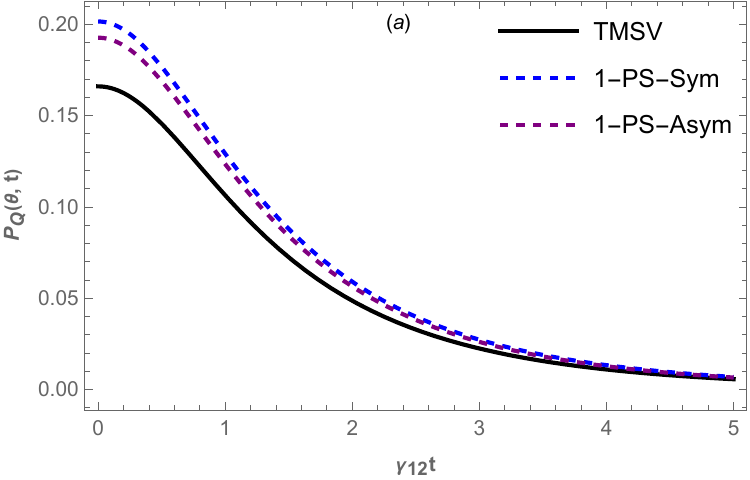}
    \includegraphics[width=\linewidth]{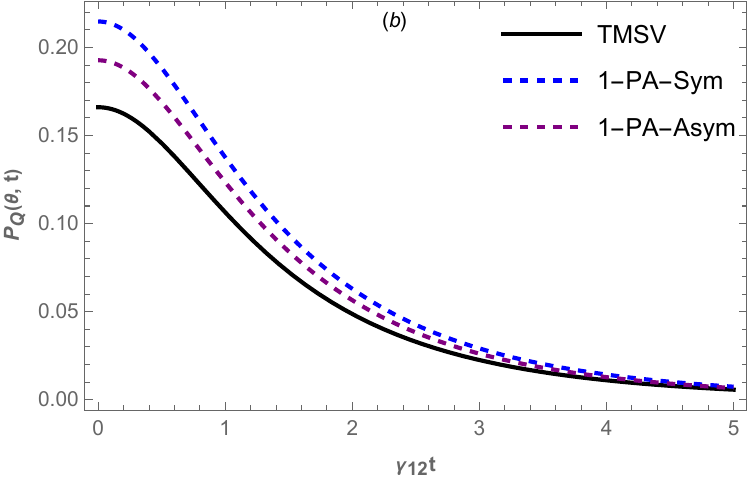}
    \includegraphics[width=\linewidth]{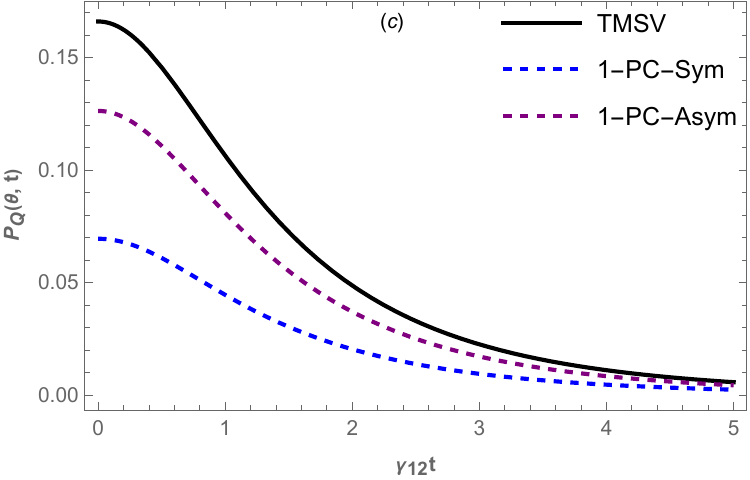}
 	\caption{   Husimi phase distribution $P_Q(\theta,t)$ as a function of scaled phase parameter $t$ for different NGTMSV  states.  We have set $\lambda=0.9$, $\tau=0.9$, $\theta_1=\theta_2=0$ and $\gamma_{1}=\gamma_{2}=\gamma_{12}$.
 	}
 	\label{ngtmsv_phase_dynamics}
 \end{figure}

The Fig.~\ref{ngtmsv_phase_dynamics} illustrates the behavior of the Husimi phase distribution \( P_Q(0,t) \) as a function of the scaled time parameter \( \gamma_{12} t\) for TMSV state and its non-Gaussian variants: photon-subtracted, photon-added, and photon-catalyzed states. The analysis is performed under the influence of an amplitude damping channel, which introduces decoherence and affects the phase distribution dynamics.

For the TMSV state (black line), the phase distribution exhibits a smooth, gradual decay as \(\gamma_{12} t \) increases, indicating a steady loss due to amplitude damping. In photon subtraction and addition cases (Fig.~\ref{ngtmsv_phase_dynamics}a-b), the initial phase distribution increases compared to TMSV, highlighting the enhanced nonclassicality of the states. The comparison between symmetric and asymmetric PA and PS cases shows minor differences, with both exhibiting a similar trend of faster decoherence compared to TMSV.
Photon catalysis (Fig.~\ref{ngtmsv_phase_dynamics}c) is seen to attain the lowest phase distribution among the non-Gaussian states. The initial \( P_Q(0,t) \) values for PC states are lower than those for PS and PA, and their decay is more gradual, suggesting that photon catalysis offers better protection against the effects of amplitude damping.\\ 
Overall, the results bring out that non-Gaussian operations significantly influence the phase distribution and robustness against decoherence. The distinction between symmetric and asymmetric operations is relatively small but observable, with asymmetric cases exhibiting slightly different decay behavior. Ultimately, amplitude damping causes all states to exhibit a monotonic decrease in phase distribution, leading them towards a more classical-like behavior over time.

\section{Result and Discussion} \label{sec: Results_Discussion}
This section presents a comprehensive analysis of how non-Gaussian operations—PS, PA and PC—affect the Husimi phase distribution of single-mode and two-mode squeezed vacuum states. Our work demonstrate clear distinctions between the operational impacts on phase localization, and robustness against amplitude damping loss/decoherence, providing a unified understanding of their role in phase-sensitive quantum technologies.

The application of non-Gaussian operations on SSV yields substantial changes in the phase distribution structure. Specifically, both PS and PA result in localization of the Husimi phase distribution, characterized by sharper peaks and reduced spread around the central phase value ($\theta=\pm \pi/2$). This localization intensifies with increase in the number of photons added or subtracted, and is further enhanced by higher squeezing ($\lambda$) and beam splitter transmittance ($\tau$). These observations are consistent with prior studies that demonstrate enhanced phase resolution and nonclassicality arising from PS and PA operations ~\cite{Agarwal92,Barnett,Nunn,malpani2019lower}.

In contrast, photon catalysis introduces delocalization, with broader and more oscillatory phase distributions. These effects suggest increased phase uncertainty, arising from quantum interference and superposition effects induced by the non-Gaussian operation. This aligns with previous findings linking photon catalysis to increased nonclassicality ~\cite{zhang2024preparation, birrittella2018photon}, suggesting its potential for quantum state engineering and metrological applications.

\begin{figure}[!ht]
    \includegraphics[width=\linewidth]{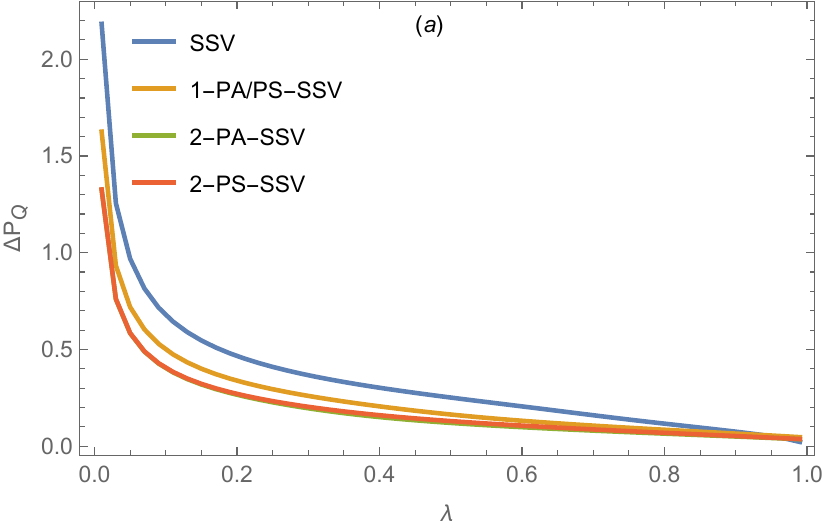}
    \includegraphics[width=\linewidth]{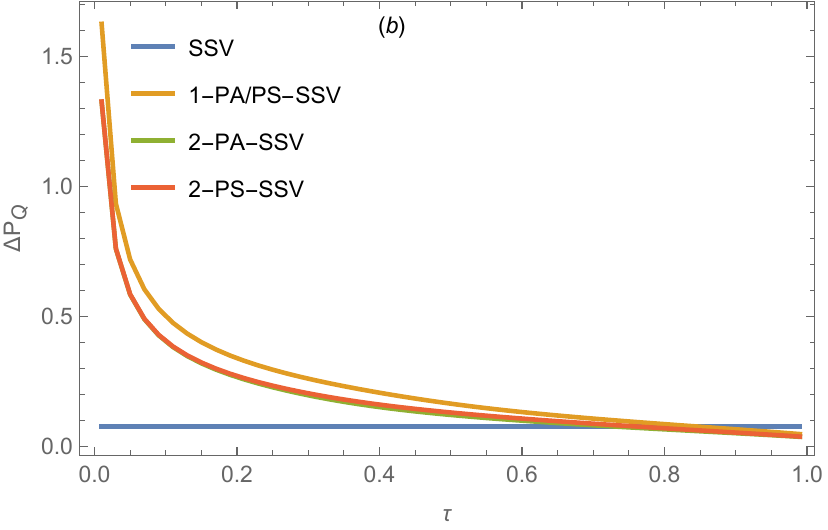}
    \caption{Standard deviation in Husimi phase distribution $P_Q(\theta)$ as a function of  a) squeezing parameter $\lambda$ and b) beam splitter transmittivity $\tau$ at center phase parameter $\theta /\pi=0.5$ for different NGSSV states.}
	\label{ngssv_phase_deviation}
\end{figure}

Quantitative analysis of the standard deviation ($\Delta P_Q$) of the phase distribution confirms these trends (Fig. \ref{ngssv_phase_deviation}). PS and PA operations lead to a monotonic decrease in $\Delta P_Q$ with increasing $\lambda$ and $\tau$, while PC states exhibit increased $\Delta P_Q$ and complex phase features (figure not shown here), suggesting their suitability in applications requiring phase sensitivity rather than precision.

Building on the insights obtained from the single-mode analysis, we extended the framework to the TMSV state. Conceptually, TMSV can be viewed as an entangled extension of two SSV states, allowing us to investigate how local non-Gaussian effects translate into non-local, correlated phase behavior. The Husimi phase distribution in this case becomes a function of the sum of the phases of the two modes, $\theta_1+\theta_2$, revealing the entanglement-induced phase locking inherent in TMSV states~\cite{PhysRevLett.112.070402}.

Our results show that the localization behavior observed in SSV states persists in the TMSV setting, with PS and PA operations sharpening the phase distribution around $\theta_1+\theta_2=0$. Importantly, symmetric operations—where the same number of photons are added or subtracted from both modes—result in enhanced phase coherence and sharper peaks compared to asymmetric cases. This reinforces the view that symmetric non-Gaussian operations act cooperatively to preserve and amplify quantum correlations in bipartite systems~\cite{PhysRevA.86.012328}.

Photon catalysis in the TMSV state, as in the SSV case, leads to broader and more intricate phase structures, although the central peak at $\theta_1+\theta_2=0$ remains intact. These results underscore the fact that non-Gaussian operations reshape, but do not erase, entanglement-induced phase correlations, offering a powerful tool for engineering joint phase distributions in continuous-variable quantum systems.

To assess the practical utility of non-Gaussian-modified squeezed states in noisy environments, we investigated their behavior under amplitude damping, a common decoherence mechanism in optical systems. Our results show that PS and PA states are more robust than their Gaussian counterparts, both in the single-mode and two-mode cases.

In the SSV system, one- and two-photon subtracted (PSSSV) and added (PASSV) states exhibit a slower decay of the Husimi phase distribution over time (Fig. \ref{ngssv_phase_dynamics}). This behavior suggests that non-Gaussianity introduced by PS and PA enhances resilience to photon loss, consistent with earlier studies on the robustness of non-Gaussian states \cite{hertz2023decoherence,meena2023characterization}.
In contrast, PC states show faster degradation, despite having higher initial sensitivity. This indicates that photon-catalyzed states are more vulnerable to decoherence, likely due to their inherently fragile interference-based phase structure.

In the TMSV system, similar dynamics was observed. The symmetric application of PS and PA operations provides greater phase stability over time compared to asymmetric implementations (Fig. \ref{ngtmsv_phase_dynamics}). These results suggest that symmetrically engineered non-Gaussian TMSV states could serve as useful resources in distributed quantum sensing and quantum metrology networks, where entangled phase coherence must be maintained over finite durations.

The insights gained from this work can be extended to multi-mode squeezed states and hybrid quantum systems, providing a broader framework for phase-sensitive quantum state engineering in practical quantum technologies.


 \section{Conclusions}\label{conc}
We investigate the Husimi phase distribution, an experimentally measurable quantity, for single-mode and two-mode squeezed states.  Our study focuses on the impact of non-Gaussian operations, i.e., photon subtraction, photon addition and photon catalysis, on the phase distribution of single-mode and two-mode squeezed vacuum states. Additionally, the impact of amplitude damping noise on the Husimi phase distribution is analyzed.
The analysis highlights that photon subtraction and photon addition operations serve as effective tools for localizing phase distribution and enhancing phase robustness in the presence of noise, while photon catalysis  enhances phase sensitivity but leads to greater delocalization. The work highlights the perspective that combined effects of squeezing, beam splitter transmittance, and environmental interactions must be carefully considered when quantum state engineering protocols are designed and phase properties provide a valuable insight into this endeavour.	

\begin{acknowledgments}
Ramniwas Meena acknowledges financial support from CSIR-HRDG through the CSIR JRF/SRF fellowship (File No. 09/1125(0015)2020-EMR-I).
\end{acknowledgments}

\bibliographystyle{unsrt}

\input{source.bbl}
\appendix
	 \begin{widetext}
\section{Matrices appearing in the Husimi distribution}\label{appwigner}

    \begin{equation}\label{mat2}
    M_2=\frac{1}{2}\left(
\begin{array}{cccc}
 \sqrt{1-\tau_1} & i \sqrt{1-\tau_1} & 0 & 0 \\
 -\sqrt{1-\tau_1} & i \sqrt{1-\tau_1} & 0 & 0 \\
 0 & 0 & \sqrt{1-\tau_2} & i \sqrt{1-\tau_2} \\
 0 & 0 & -\sqrt{1-\tau_2} & i \sqrt{1-\tau_2} \\
 0 & 0 & -\lambda  \sqrt{1-\tau_1} \sqrt{\tau_2} & i \lambda  \sqrt{1-\tau_1} \sqrt{\tau_2} \\
 0 & 0 & \lambda  \sqrt{1-\tau_1} \sqrt{\tau_2} & i \lambda  \sqrt{1-\tau_1} \sqrt{\tau_2} \\
 -\lambda  \sqrt{\tau_1} \sqrt{1-\tau_2} & i \lambda  \sqrt{\tau_1} \sqrt{1-\tau_2} & 0 & 0 \\
 \lambda  \sqrt{\tau_1} \sqrt{1-\tau_2} & i \lambda  \sqrt{\tau_1} \sqrt{1-\tau_2} & 0 & 0 \\
\end{array}
\right),
\end{equation}

\begin{equation}\label{mat3}
    M_3=\frac{-1}{4}\left(
\begin{array}{cccccccc}
 0 & 0 & 0 & 0 & 0 & \sqrt{\tau_1} & 0 & 0 \\
 0 & 0 & 0 & 0 & \sqrt{\tau_1} & 0 & 0 & 0 \\
 0 & 0 & 0 & 0 & 0 & 0 & 0 & \sqrt{\tau_2} \\
 0 & 0 & 0 & 0 & 0 & 0 & \sqrt{\tau_2} & 0 \\
 0 & \sqrt{\tau_1} & 0 & 0 & 0 & 0 & -\lambda  \sqrt{1-\tau_1} \sqrt{1-\tau_2} & 0 \\
 \sqrt{\tau_1} & 0 & 0 & 0 & 0 & 0 & 0 & -\lambda  \sqrt{1-\tau_1} \sqrt{1-\tau_2} \\
 0 & 0 & 0 & \sqrt{\tau_2} & -\lambda  \sqrt{1-\tau_1} \sqrt{1-\tau_2} & 0 & 0 & 0 \\
 0 & 0 & \sqrt{\tau_2} & 0 & 0 & -\lambda  \sqrt{1-\tau_1} \sqrt{1-\tau_2} & 0 & 0 \\
\end{array}
\right),
\end{equation}


\begin{equation}\label{mat4}
	M_4=\frac{1}{4-4 \lambda ^2 \tau_1 \tau_2} \left(
	\begin{array}{cccccccc}
		0 & c_1 & c_2 & 0 & 0 & c_3 & c_4 & 0 \\
		c_1 & 0 & 0 & c_2 & c_3 & 0 & 0 & c_4 \\
		c_2 & 0 & 0 & c_5 & c_6 & 0 & 0 & c_7 \\
		0 & c_2 & c_5 & 0 & 0 & c_6 & c_7 & 0 \\
		0 & c_3 & c_6 & 0 & 0 & c_8 & c_9 & 0 \\
		c_3 & 0 & 0 & c_6 & c_8 & 0 & 0 & c_9 \\
		c_4 & 0 & 0 & c_7 & c_9 & 0 & 0 & c_{10} \\
		0 & c_4 & c_7 & 0 & 0 & c_9 & c_{10} & 0 \\
	\end{array}
	\right),
\end{equation}
with the coefficients $c_i$'s being
\begin{equation}
	\left(
	\begin{array}{ccc}
		c_1 & = & \tau_1-1 \\
		c_2 & = & \lambda  \sqrt{(\tau_1-1) \tau_1 (\tau_2-1) \tau_2} \\
		c_3 & = & \sqrt{\tau_1} \left(\lambda ^2 \tau_2-1\right) \\
		c_4 & = & -\lambda  \sqrt{(\tau_1-1) \tau_1 (\tau_2-1)} \\
		c_5 & = & \tau_2-1 \\
		c_6 & = & -\lambda  \sqrt{(\tau_1-1) (\tau_2-1) \tau_2} \\
		c_7 & = & \sqrt{\tau_2} \left(\lambda ^2 \tau_1-1\right) \\
		c_8 & = & \lambda ^2 (\tau_1-1) \tau_2 \\
		c_9 & = & \lambda  \sqrt{(\tau_1-1) (\tau_2-1)} \\
		c_{10} & = & \lambda ^2 \tau_1 (\tau_2-1) \\
	\end{array}
	\right).
\end{equation}
\end{widetext}

\section{Solution of the Equation ~\eqref{eq:multimaster_sol_term}}
We define the following super-operators: \(\hat{A}\cdot = \hat{a}_i \cdot \hat{a}_i^\dag\), \(\hat{B}\cdot = \hat{a}_i^\dag \hat{a}_i \cdot\), and \(\hat{C}\cdot = \cdot \hat{a}_i^\dag \hat{a}_i\), which operate on the density matrix \(\rho\). Specifically, \(\hat{A}\rho = \hat{a}_i \rho \hat{a}_i^\dag\), \(\hat{B}\rho = \hat{a}_i^\dag \hat{a}_i \rho\), and \(\hat{C}\rho = \rho \hat{a}_i^\dag \hat{a}_i\). Using the commutator relation \([\hat{a}_i, \hat{a}_i^\dag] = 1\), we can deduce the commutation relations for these super-operators:
\[
    [\hat{A}, \hat{B}] = [\hat{A}, \hat{C}] = \hat{A}, \quad [\hat{B}, \hat{C}] = 0.
\]
Next, by introducing \(\hat{D} = 2\hat{A} - \hat{B} - \hat{C}\), we express the left part of  Eq. \eqref{eq:multimaster_sol_term} as:
\[
    e^{\frac{\gamma_i}{2}   L_i} \rho(0) = e^{\frac{\gamma_i }{2}\hat{D}} \rho(0),
\]
where \(\gamma_i\) is  damping rate of $i^{th}$ mode.

Further simplification yields:
\[ e^{\frac{\gamma_i t}{2} \hat{D}} = e^{\frac{\gamma_i t}{2} (2\hat{A} - \hat{B} - \hat{C})} = e^{[e^{{\gamma_i t}} - 1] \hat{A}} e^{-\frac{\gamma_i t}{2} (\hat{B} + \hat{C})}. \]

Here, the Zassenhaus formula \cite{dupays2023closed} and the commutation relations have been applied, using the identity (if $[\hat{X},\hat{Y}]=x \hat{X}$, where  $x$ is a constant):
\begin{equation}
    \exp\left[ a \hat{X} + b  \hat{Y}\right] = e^{-\frac{a}{bx}[e^{-bx} - 1] \hat{X}} e^{b \hat{Y}}\quad,
\end{equation} 
with \(\hat{X} = 2\hat{A}\), \(\hat{Y} = \hat{B} + \hat{C}\),  \(a=-b = \frac{\gamma_i t}{2}\) and $[\hat{X}, \hat{Y}]=2\hat{X}$.

Moreover, the terms can be simplified as:
\[ e^{-\frac{\gamma_i t}{2} (\hat{B} + \hat{C})} = e^{-\frac{\gamma_i t}{2} \hat{B}} e^{-\frac{\gamma_i t}{2} \hat{C}}. \]

Thus, the solution becomes:
\begin{align}
     e^{\frac{\gamma_i t}{2} \hat{D}} \rho(0)&= e^{[e^{{\gamma_i t}} - 1] \hat{A}} e^{-\frac{\gamma_i t}{2} \hat{B}} e^{-\frac{\gamma_i t}{2} \hat{C}} \rho(0)\nonumber \\
    &= e^{[e^{{\gamma_i t}} - 1] \hat{A}} e^{-\frac{\gamma_i t}{2} \hat{a}_i^\dag \hat{a}_i} \rho(0) e^{-\frac{\gamma_i t}{2} \hat{a}_i^\dag \hat{a}_i}\nonumber \\
    &= \sum_{n=0}^{\infty} \left( \frac{[e^{{\gamma_i t}} - 1]^{n}}{n!} \hat{a}_i^{n} e^{-\frac{\gamma_i t}{2} \hat{a}_i^\dag \hat{a}_i} \rho(0) e^{-\frac{\gamma_i t}{2} \hat{a}_i^\dag \hat{a}_i} \hat{a}_i^{\dag n} \right).
\end{align}

For an arbitrary initial element \( \rho(0) = |\alpha\rangle\langle\alpha| \), its time evolution is given by:
\begin{align}
    (|\alpha\rangle\langle\alpha|)_t = \sum_{n=0}^{\infty} \frac{[e^{{\gamma_i t}} - 1]^{n}}{n!} \hat{a}_i^n \exp\left(-\frac{\gamma_i t}{2} \hat{a}_i^\dag \hat{a}_i \right) \nonumber\\
     |\alpha\rangle \langle \alpha| \exp\left(-\frac{\gamma_i t}{2}  \hat{a}_i^\dag \hat{a}_i \right) \hat{a}_i^{\dag n}.
\end{align}

To handle the action of the operator \( \exp\left(-\frac{\gamma_i t}{2}  \hat{a}_i^\dag \hat{a}_i\right) \) on the coherent state \( |\alpha\rangle \), we begin by considering its effect on the \( |\alpha\rangle \):
\begin{align}
    &\exp\left(-\frac{\gamma_i t}{2}  \hat{a}_i^\dag \hat{a}_i \right)|\alpha\rangle \nonumber \\
    &=\exp\left(-\frac{|\alpha|^2}{2}\right) \sum_{n=0}^{\infty} \frac{\alpha^n}{\sqrt{n!}}  \exp\left(-\frac{\gamma_i t}{2}  \hat{a}_i^\dag \hat{a}_i \right) |n\rangle \nonumber \\
    &=\exp\left(-\frac{|\alpha|^2}{2}\right) \sum_{n=0}^{\infty} \frac{\alpha^n}{\sqrt{n!}}  \exp\left(-\frac{ n \gamma_i t}{2}   \right) |n\rangle \nonumber \\
    &=\exp\left(-\frac{|\alpha|^2}{2}\right) \sum_{n=0}^{\infty} \frac{{\left(\alpha  e^{-\frac{\gamma_i t}{2}  }\right)}^n}{\sqrt{n!}}  |n\rangle \nonumber \\
    &=\exp\left(-\frac{|\alpha|^2}{2} \left[1- e^{-\gamma_i t} \right]\right)  |\alpha  e^{-\frac{\gamma_i t}{2} }\rangle
\end{align}
Similarly, for \( \langle \alpha| \), we have

\begin{align}
    \langle\alpha|\exp\left(-\frac{\gamma_i t}{2}  \hat{a}_i^\dag \hat{a}_i \right) \nonumber =\langle\alpha  e^{-\frac{\gamma_i t}{2} }|\exp\left(-\frac{|\alpha|^2}{2} \left[1- e^{-\gamma_i t} \right]\right).  
\end{align}

Combining these, we can now express the time-evolved state \( (|\alpha\rangle \langle \alpha|)_t \) as:
\begin{widetext}
    \begin{align}
    (|\alpha\rangle\langle\alpha|)_t \nonumber 
    &= \sum_{n=0}^{\infty} \frac{[e^{{\gamma_i t}} - 1]^{n}}{n!} \hat{a}_i^n  \exp\left(-{|\alpha|^2} \left[1- e^{-\gamma_i t} \right]\right)  |\alpha  e^{-\frac{\gamma_i t}{2} }\rangle \langle\alpha  e^{-\frac{\gamma_i t}{2} }|
      \hat{a}_i^{\dag n} \nonumber \\
    &= \sum_{n=0}^{\infty} \frac{[e^{{\gamma_i t}} - 1]^{n}}{n!} \left(|\alpha|^2  e^{-{\gamma_i t} }\right)^n  \exp\left(-{|\alpha|^2} \left[1- e^{-\gamma_i t} \right]\right)  |\alpha  e^{-\frac{\gamma_i t}{2} }\rangle \langle\alpha  e^{-\frac{\gamma_i t}{2} }| \nonumber\\
    &=   \exp\left(-{|\alpha|^2} \left[1- e^{-\gamma_i t} \right]\right) \sum_{n=0}^{\infty} \frac{[e^{{\gamma_i t}} - 1]^{n}}{n!} \left(|\alpha|^2  e^{-\gamma_i t }\right)^n |\alpha  e^{-\frac{\gamma_i t}{2} }\rangle \langle\alpha  e^{-\frac{\gamma_i t}{2} }|\nonumber\\
    &=   \exp\left(-{|\alpha|^2} \left[1- e^{-\gamma_i t} \right]\right) \exp\left({|\alpha|^2} \left[1- e^{-\gamma_i t} \right]\right)  |\alpha  e^{-\frac{\gamma_i t}{2} }\rangle \langle\alpha  e^{-\frac{\gamma_i t}{2} }|\nonumber\\
    &= |\alpha  e^{-\frac{\gamma_i t}{2} }\rangle \langle\alpha  e^{-\frac{\gamma_i t}{2} }|.
\end{align}
\end{widetext}

\section{Evolution of Husimi $Q$-function}
We can express the density operator of the non-Gaussian squeezed state vacuum (NGSSV) in terms of the coherent state basis using the P-function representation as follows:
\begin{equation}
    \rho_{NG}=\int{d^2\alpha \;P(\alpha)|\alpha\rangle\langle\alpha|} \label{Input_NGSSV}
\end{equation}
where the density matrix for NGSSV evolves over time according to the expression:
\begin{equation}
     \rho_{NG}(t)=\int{d^2\alpha \; P(\alpha)|\alpha e^{-\frac{\gamma_1 t}{2}}\rangle\langle\alpha e^{-\frac{\gamma_1 t}{2}}|}.  \label{output_NGSSV}
\end{equation}
Next, the Husimi characteristic function (in Anti-normal ordering) can be written as \cite{gerry2023introductory}:
\begin{align}
    &\chi_{AN}(\Lambda)\nonumber \\&=Tr\left[\hat\rho \exp(\Lambda \hat{a}^\dag -\Lambda^{*} \hat{a}-(1/2) |\Lambda|^2\right]\nonumber \\
    &=\frac{1}{\pi}\int{d^2\beta \langle \beta | e^{-\Lambda \hat{a}^\dag}\rho_{NG} e ^{\Lambda^{*}\hat{a}}|\beta\rangle} \nonumber \\
    &=\frac{1}{\pi}\int{d^2\beta \langle \beta | e^{-\Lambda \hat{a}^\dag}\int{d^2\alpha P(\alpha)|\alpha\rangle\langle\alpha|} e ^{\Lambda^{*}\hat{a}}|\beta\rangle}  \nonumber \\
    &=\frac{1}{\pi}\int\int{d^2\beta d^2\alpha P(\alpha) e^{\Lambda^{*} \beta-\Lambda \beta^{*}}   |\langle \beta |\alpha\rangle|^2} \nonumber \\
    &=\frac{1}{\pi}\int\int{d^2\beta d^2\alpha P(\alpha) e^{\Lambda^{*} \beta-\Lambda \beta^{*}}  e^{- |\beta -\alpha|^2}} \nonumber \\
    &=\frac{1}{\pi}\int\int{d^2\beta d^2\alpha P(\alpha) e^{\Lambda^{*} \beta-\Lambda \beta^{*}}  e^{- |\beta|^2 -|\alpha|^2+\alpha \beta^{*}+\alpha^{*} \beta}} \nonumber \\
    &=\frac{1}{\pi}\int\int{d^2\beta d^2\alpha P(\alpha)   e^{- |\beta|^2 -|\alpha|^2-(\Lambda-\alpha) \beta^{*}+(\Lambda^{*}+\alpha^{*} )\beta}} \nonumber
\end{align}
By using the following result:
\begin{equation}
    \int{\frac{d^2 \gamma}{\pi} e^{\zeta |\gamma|^2+\xi \gamma+\eta \gamma^{*}}}=-\frac{1}{\zeta} e^{-\left(\xi \eta/\zeta\right)}\quad \text{if } \text{Re}\;\zeta<0, \label{Eq:int_formula}
\end{equation}
we get
\begin{align}
    \chi_{AN}(\Lambda)\nonumber \
    &=\int{ d^2\alpha P(\alpha)   e^{-|\alpha|^2-(\Lambda-\alpha) (\Lambda^{*}+\alpha^{*} )}} \nonumber \\
    &=\int{ d^2\alpha P(\alpha)   e^{-|\Lambda|^2-\Lambda\alpha^{*}+ \Lambda^{*}\alpha }} \label{CF_AN}.
\end{align}
The evolution of the Husimi characteristic function over time is then given by:
\begin{align}
    \chi_{AN}(\Lambda,t)&=\frac{1}{\pi}\int{d^2\beta \langle \beta | e^{-\Lambda \hat{a}^\dag}\rho_{NG}(t) e ^{\Lambda^{*}\hat{a}}|\beta\rangle} \nonumber \\
    &=\int{ d^2\alpha P(\alpha)   e^{-|\Lambda|^2-\Lambda\alpha^{*}e^{-\frac{\gamma_1 t}{2}}+ \Lambda^{*}\alpha e^{-\frac{\gamma_1 t}{2}} }} \nonumber \\
    &= e^{-|\Lambda|^2+|\bar{\Lambda}|^2}\int{ d^2\alpha P(\alpha)   e^{-|\bar{\Lambda}|^2-\bar{\Lambda}\alpha^{*}+ \bar{\Lambda}^{*}\alpha }}, \label{CF_AN_t0}
\end{align}
where $\bar{\Lambda}=\Lambda e^{-\frac{\gamma_1 t}{2}}$.
Now using Eq. \eqref{CF_AN} and \eqref{CF_AN_t0}, 
\begin{align}
    \chi_{AN}(\Lambda,t)&= e^{-T|\Lambda|^2 } \chi_{AN}\left(\Lambda e^{-\frac{\gamma_1 t}{2}}\right);\;\;\; T=\left(1-e^{-\gamma_1 t}\right) \label{CF_AN_t}.  
\end{align}

The time-evolved Husimi \(Q\)-function is given by \cite{gerry2023introductory}
\begin{align}
    &Q(\alpha,t)\nonumber\\
    &=\int{\frac{d^2 \Lambda}{\pi^2} e^{\alpha \Lambda^{*}-\alpha^{*}\Lambda} \chi_A(\Lambda,t)}\nonumber \\
    &=\int{\frac{d^2 \Lambda}{\pi^2}  \exp{\left[- T|\Lambda|^2+\alpha \Lambda^{*}-\alpha^{*}\Lambda\right]\chi_{A}(\Lambda e^{-\frac{\gamma_1 t}{2}})}}\nonumber \\
    &=e^{\gamma_1 t}\int\frac{d^2 \bar{\Lambda}}{\pi^2}  \exp{\left[-|\bar{\Lambda}|^2\left(e^{\gamma_1 t}-1\right)+\alpha \bar{\Lambda}^{*}e^{\frac{\gamma_1 t}{2}}-\alpha^{*}\bar{\Lambda}e^{\frac{\gamma_1 t}{2}}\right]}\nonumber \\
    &\times\chi_{A}(\bar{\Lambda}). \nonumber
\end{align}
By using the  relation, 
\begin{equation}
    \chi_{A}(\bar{\Lambda})=\int{e^{\bar{\Lambda}\beta^*-\bar{\Lambda}^{*}\beta}Q(\beta) d^2\beta},
\end{equation}
we get, 
\begin{align}
    &Q(\alpha,t)\nonumber\\
    &=e^{\gamma_1 t}\int\int \exp{\left[-|\bar{\Lambda}|^2\left(e^{\gamma_1 t}-1\right)+\alpha \bar{\Lambda}^{*}e^{\frac{\gamma_1 t}{2}}-\alpha^{*}\bar{\Lambda}e^{\frac{\gamma_1 t}{2}}\right]}\nonumber \\
    &\times e^{\bar{\Lambda} \beta^{*}-\bar{\Lambda}^{*}\beta}Q(\beta) \frac{d^2 \bar{\Lambda}}{\pi}\frac{d^2\beta}{\pi}. \label{Qt_NGSSV1}
\end{align}
Performing the integration over $\bar\Lambda$ (using Eq. \eqref{Eq:int_formula}), this simplifies to
\begin{equation}
 Q(\alpha, t) = \int \frac{d^2 \beta}{\pi \left( 1 - e^{-\gamma_1 t} \right)} \exp\left[ -\frac{|\alpha - \beta e^{-\frac{\gamma_1 t}{2}}|^2}{1 - e^{-\gamma_1 t}} \right] Q(\beta). \label{Qt_NGSSV}   
\end{equation}
\begin{widetext}
Similarly, the time-evolved \(Q\)-function for two-mode can be written as
\begin{multline}
 Q(\alpha_1,\alpha_2, t) = \int \frac{d^2 \beta_1 d^2 \beta_2}{\pi^2 \left( 1 - e^{-\gamma_1 t} \right) \left( 1 - e^{-\gamma_2 t} \right)}  \exp\left[ -\frac{|\alpha_1 - \beta_1 e^{-\frac{\gamma_1 t}{2}}|^2 }{\left(1 - e^{-\gamma_1 t}\right)} -\frac{|\alpha_2 - \beta_2 e^{-\frac{\gamma_2 t}{2}}|^2}{\left(1 - e^{-\gamma_2 t}\right)} \right] Q(\beta_1,\beta_2). \label{Qt_NGTMSV} 
\end{multline}

\end{widetext}

\end{document}

%% file: source.bbl
%